\documentclass[preprint]{elsarticle}

\usepackage{bbm}
\usepackage{lineno,hyperref}
\modulolinenumbers[5]

\journal{Progress in Quantum Electronics}

\begin{document}

\begin{frontmatter}

%\title{Photon-by-photon quantum light state engineering\\ Engineering light photon-by-photon\\ Engineering the quantum state of light one photon at a time}

\title{Photon-by-photon quantum light state engineering}

%%% Group authors per affiliation:

\author[mainaddress,secondaryaddress]{Nicola Biagi}
\author[mainaddress,secondaryaddress]{Saverio Francesconi}
\author[mainaddress,secondaryaddress]{Alessandro Zavatta}
\author[mainaddress,secondaryaddress]{Marco Bellini\corref{mycorrespondingauthor}}

\cortext[mycorrespondingauthor]{Corresponding author}
\ead{marco.bellini@ino.cnr.it}

\address[mainaddress]{Istituto Nazionale di Ottica (CNR-INO), L.go E. Fermi 6, 50125 Florence, Italy}
\address[secondaryaddress]{LENS and Department of Physics $\&$ Astronomy, University of Firenze, 50019 Sesto Fiorentino, Florence, Italy}

\begin{abstract}
The ability to manipulate light at the level of single photons, its elementary excitation quanta, has recently made it possible to produce a rich variety of tailor-made quantum states and arbitrary quantum operations, of high interest for fundamental science and applications. Here we present a concise review of the progress made over the last few decades in the engineering of quantum light states. Although far from exhaustive, this review aims at providing a sufficiently wide and updated introduction that may serve as the entry point to such a fascinating and rapidly evolving field.
\end{abstract}

\begin{keyword}
Quantum Optics, Photons, Quantum state engineering
\end{keyword}

\end{frontmatter}

%\linenumbers

\section{Introduction}
The possibility of controlling light at the most intimate level is a fundamental requisite for studying its quantum nature and exploiting it in applications towards future technologies. The manipulation of single quanta of the electromagnetic field in the optical domain is a relatively recent achievement, but it has already allowed researchers all over the world to perform experiments that investigate the very foundations of quantum mechanics \cite{parigi_07_probinga} and pose the bases of novel quantum-enhanced technologies.

Individual photons can be generated in well-defined field modes either deterministically or in a probabilistic heralded fashion with high purity and high rates, and single photons may also be added to \cite{zavatta_04_quantumtoclassical} and subtracted from \cite{wenger_04_nongaussian} arbitrary light states one by one, in order to precisely sculpt light with the desired properties. When acting on single, well-defined modes, these simple operations are described by the bosonic creation and annihilation operators $\hat a^{\dag}$ and $\hat a$. In recent years, besides applying these two operations separately, it also became possible to combine them in sequences \cite{zavatta_11_highfidelity} and coherent superpositions \cite{zavatta_09_experimental} to further expand the range of available manipulation tools.

Even more recently, it has been shown how to perform these operations not just on a single mode of the electromagnetic field, but coherently over two or more modes \cite{ourjoumtsev_07_increasing,biagi_21_coherent}, opening the possibility to generate and accurately control the entanglement in a general multimode system. Various types of modes have been used: from separate spatial modes in the form of different light beams, to distinct traveling temporal wavepackets, or sets of ultrafast orthogonal time-frequency modes. 

In this review we will concisely describe recent advances in the engineering of optical quantum light states by means of the operations of photon addition and subtraction, in single and multiple distinct modes. We will concentrate on the most used approach to non-deterministically implement photon addition, that is by means of parametric down-conversion in a nonlinear crystal, and on photon subtraction by a low-reflectivity beam-splitter, but we will also mention novel schemes, for example those using sum-frequency generation for subtracting photons in arbitrary time-frequency modes \cite{ra_17_tomography}. 
We will start by introducing the basic tools of quantum light state engineering and then move from their use in single-mode state manipulation to their application in multimode scenarios.

Of course, this overview is far from exhaustive and it is mainly focused on the use of the non-Gaussian operations of photon addition and subtraction for engineering the quantum state of light. Although previous review papers introducing these topics had already appeared more than one decade ago \cite{kim_08_recent,bellini_10_manipulating}, this work aims at presenting a much wider and more updated description of current research.
For a good introduction to the basic ingredients of quantum optics and to the generation, manipulation and characterization of optical quantum states, one may refer to the book of Leonhardt \cite{leonhardt_97_measuring} or to the brief tutorial by Olivares \cite{olivares_21_introduction}. The analysis of continuous-variable quantum states of light by homodyne detection and quantum tomography is also well reviewed in \cite{lvovsky_09_continuousvariable}. 
A closer look at quantum states of light and at their multimode description can be found in \cite{fabre_20_modes}, while a concise illustration of the different states produced by coherent multimode photon addition is presented in \cite{biagi_21_coherent}.
Recent reviews on the general field of non-Gaussian states, from their description to their production and applications can be found in \cite{walschaers_21_nongaussiana} and \cite{lvovsky_20_production}. Finally, the interested reader may learn more about the specific topic of hybrid continuous/discrete-variable approaches to quantum information in \cite{vanloock_11_optical} and \cite{andersen_15_hybrid}.

\section{Single-mode quantum state engineering}

Any pure single-mode quantum state of a field can be written as a weighted superposition of Fock states $\vert n \rangle$, the eigenstates of the free Hamiltonian of the field, in the following form: 
\begin{equation}
	\vert \psi \rangle=\sum_{n=0}^{\infty}c_n \vert n \rangle
	\label{eq_fock_sup}
\end{equation}
with $\sum|c_n|^2=1$.
All the properties of the field are solely determined by the complex coefficients $c_n$, and quantum state engineering is thus the art of manipulating them in order to generate arbitrary states of light.
Producing the general superposition of Fock states of Eq.(\ref{eq_fock_sup}) is nontrivial, as it would require highly nonlinear photon-photon interactions, which are not normally available in nature. Therefore, most of the recent spectacular achievements in photonic quantum state engineering have been based on alternative schemes using readily available resources, such as beam-splitters, parametric down-converters, homodyne detectors and photon counters.

\subsection{The basic tools of quantum light state engineering}

The main building blocks for the implementation of many quantum state engineering tasks are two simple devices, common in any optical laboratory: the beam-splitter (BS) and the parametric down-converter (PDC). Although apparently quite different, they share a very similar quantum description, and are often used together to produce complex quantum operations. One can think of them as devices with two input and two output ports.

Let us start by considering the beam-splitter operator \cite{gerry_04_introductory}
\begin{equation}
	\hat{B}(\tau)=\exp \left[ \tau (\hat{a}\hat{b}^{\dagger}-\hat{a}^{\dagger}\hat{b})\right],
	\label{eq_bs}
\end{equation}
where $\hat{a}$ and $\hat{b}$ ($\hat{a}^{\dagger}$ and $\hat{b}^{\dagger}$) are bosonic annihilation (creation) operators for the BS input modes, and $\tau$ is a parameter that determines its amplitude transmittivity $t=\cos \tau$ and reflectivity $r=\sin \tau$. When $\tau$ is small, i.e. $r \ll 1$, using the Taylor expansion the BS operator can be approximated as
\begin{equation}
	\hat{B}(\tau) \approx 1 + \tau (\hat{a}\hat{b}^{\dagger}-\hat{a}^{\dagger}\hat{b}).
	\label{eq_bs_approx}
\end{equation}

Instead of beam splitters, which are essentially passive devices, one can also use an active device, which needs extra energy to operate. The parametric down-converter is one of such devices, made of a nonlinear crystal that converts a pump incident photon into two daughter photons, while conserving energy and momentum. Its operation is described by a (two-mode) squeezing operator \cite{gerry_04_introductory}
\begin{equation}
	\hat{S}(\zeta)=\exp \left[ \zeta (\hat{a}^{\dagger}\hat{b}^{\dagger}-\hat{a}\hat{b})\right].
	\label{eq_pdc}
\end{equation}
As it is often the case, for small squeezing coefficients $\zeta \ll 1$, the PDC operation can be approximated as
\begin{equation}
	\hat{S}(\zeta) \approx 1 + \zeta (\hat{a}^{\dagger}\hat{b}^{\dagger}-\hat{a}\hat{b}).
	\label{eq_pdc_approx}
\end{equation}

\begin{figure}[h]
	\centering
	\includegraphics[width=12 cm]{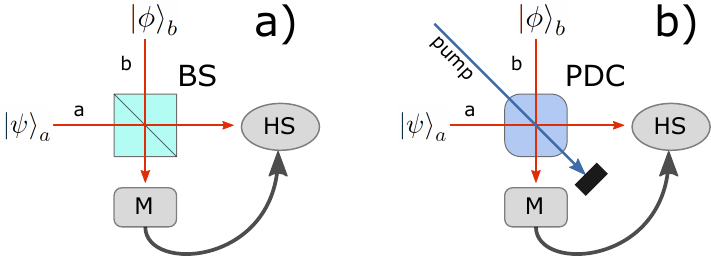}
	\caption{a) Beam-splitter (BS) and b) parametric down-converter (PDC). Conditional state engineering of the input state $\vert \psi \rangle_a$ in mode $a$ can be achieved by injecting an ancilla state $\vert \phi \rangle_b$ and performing a measurement (M) in the output mode $b$. A particular result of such a measurement heralds the generation of the heralded state (HS) in the output mode $a$.  \label{fig_BS_PDC}}
\end{figure}

If an arbitrary light state and an ancillary one are sent into the two inputs, the light state at one of the outputs can be remotely engineered conditioned on the result of different measurements performed on the ancillary output port. (see Figs.\ref{fig_BS_PDC}).

Typical measurements that one may perform are balanced homodyne detection and photon counting.

\begin{figure}[h]
	\centering
	\includegraphics[width=6 cm]{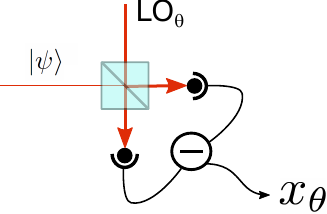}
	\caption{Homodyne detection (HD) setup. The investigated light state is mixed with an intense reference local oscillator (LO$_{\theta}$) field on a 50$\%$ beam-splitter and the two outputs are detected by proportional photodiodes. The difference photocurrent is proportional to the quadrature of the input state $\vert \psi \rangle$ at the phase $\theta$ of the LO.\label{fig_HD}}
\end{figure}

Balanced homodyne detection (HD) allows one to measure the phase-dependent electromagnetic field quadrature
\begin{equation}
	\hat{x}_{\theta} = \frac{\hat{a}^{\dagger} e^{i \theta}+\hat{a} e^{-i \theta}}{\sqrt{2}} 
	\label{eq_quad}
\end{equation}
by mixing the field of interest with an intense reference coherent field (named Local Oscillator, LO) on a 50$\%$ BS and taking the difference photocurrent signal at the two outputs as shown in Fig.\ref{fig_HD}.
Acquiring the statistics of such HD measurements for various values of the phase $\theta$ also enables one to perform quantum state tomography for reconstructing the density matrix and the Wigner function of the state \cite{leonhardt_97_measuring,lvovsky_09_continuousvariable}.

Only using beam-splitters or parametric down-converters, combined with the conditioning implemented by homodyne measurements, limits the range of the possible state engineering operations to the so-called Gaussian regime. Gaussian states are characterized by quadrature probability distributions with Gaussian shapes and are the most frequently occurring states in nature (the vacuum, coherent, and thermal states are all Gaussian, together with the nonclassical squeezed states). However, getting out of the Gaussian domain is essential for some nontrivial quantum information processing and communication task \cite{eisert_02_distilling,fiurasek_02_gaussian,giedke_02_characterization}, therefore non-Gaussian operations have to be added to our toolbox.

Photon counting is the simplest non-Gaussian measurement in quantum optics, ideally projecting the measured mode onto a non-Gaussian $ n $-photon Fock state $\vert n \rangle$.
Although intrinsically photon-number-resolving detectors (PNRDs) exist and keep attracting tremendous research interest, efficient, fast, and low-noise sensors with large enough counting capabilities (the maximum resolvable number of photons is currently limited to 4-7, depending on the platform) are not so widely available yet. Furthermore, both the transition-edge \cite{lita_08_counting,fukuda_11_titaniumbased} and the superconducting-nanowire \cite{cahall_17_multiphoton,zhu_20_resolving} sensors typically used at the core of PNRDs need to be operated at cryogenic temperatures, which also significantly limits their widespread use.

One thus often resorts to simple and readily available avalanche photodiodes (APDs) operating in the Geiger (also called on/off) mode, where one cannot discriminate the number of photons but only discern their presence or absence. A rough approximation of a PNRD can be achieved with multiple on/off APDs, by multiplexing the optical field into separate spatial or temporal channels and counting coincident detection events \cite{fitch_03_photonnumber,achilles_03_fiberassisted,achilles_06_direct,laiho_10_probing}. Of course, the probability that two or more photons reach the same detector has to be made negligible, therefore the number of detectors normally has to be larger than the number of photons to detect.

\subsection{Fock state preparation}

When the parametric down-converter device is used without injecting any field in the input modes, a spontaneous PDC process takes place and pump photons are annihilated to give birth to pairs of down-converted photons obeying energy and momentum conservation. Detecting one photon in the output mode $ b $ heralds the presence of another photon in the conjugated output mode $ a $. The operation is simply described in the low-gain approximation by:
\begin{equation}
		_b \langle 1 \vert \hat{S}(\zeta) \vert 0 \rangle _a \vert 0 \rangle _b \approx \; _b \langle 1 \vert 1 + \zeta (\hat{a}^{\dagger}\hat{b}^{\dagger}-\hat{a}\hat{b}) \vert 0 \rangle _a \vert 0 \rangle _b \rightarrow \vert 1 \rangle _a.
		\label{eq_fock}
\end{equation}

The heralded generation and the full quantum tomographic characterization of single-photon Fock states was first achieved in 2001, with the reconstruction of a negative Wigner function, one of the strongest evidences of a nonclassical character \cite{lvovsky_01_quantum}. Note that, in order to obtain spatially and spectrally pure heralded single photons, one generally needs to place narrow spatial and spectral filters with widths much smaller than the pump bandwidths in the idler path \cite{aichele_02_optical}.
Since the first heralded generation of a single photon, many experiments have followed, both in the pulsed and in the continuous-wave (CW) regime \cite{zavatta_04_tomographic,neergaard-nielsen_07_high,morin_12_highfidelity}. 

If the squeezing parameter is not negligibly small, the lowest-order approximation is no longer valid and higher terms have to be included in the expansion of the PDC operator. This means that a higher, but strictly correlated, number of photons can be spontaneously emitted in the two outputs of the crystal. The generated state is a so-called two-mode squeezed or EPR state, which can be written as
\begin{equation}
	\vert \psi _{EPR}\rangle=\hat{S}(\zeta) \vert 0 \rangle _a \vert 0 \rangle _b = \sqrt{1- \lambda^2}\sum_{n=0}^{\infty} \lambda^n \vert n \rangle _a \vert n \rangle _b
	\label{eq_epr}
\end{equation}
where $ \lambda=\tanh \zeta $
\cite{ourjoumtsev_06_quantum,zavatta_08_quantum}.
\begin{figure}[h]
	\centering
	\includegraphics[width=7 cm]{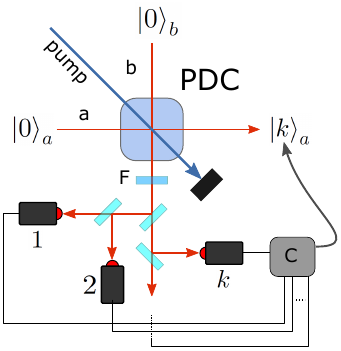}
	\caption{Scheme for heralded Fock state $ \vert k \rangle $ generation by multiple coincident clicks from $ k $ on/off detectors after spontaneous parametric down-conversion. \label{fig_fock}}
\end{figure}
If precisely $ k $ photons are detected in one mode, then a $ k $-photon Fock state is conditionally generated in the other, even though the success rate decreases exponentially with the photon number. To overcome the scarce availability of true photon-number resolving detectors, beam-splitters may be used to split the heralding field towards two or more APDs (see Fig. \ref{fig_fock}). A coincident detection of $ k $ events reveals that at least $ k $ photons have been produced and, if higher-order terms are negligible, a $ k $-photon Fock state $ \vert k \rangle $ has been generated. This approach was adopted for producing and tomographically reconstruct heralded two-photon Fock states $ \vert 2 \rangle $ \cite{ourjoumtsev_06_quantum,zavatta_08_quantum}. 
Conditioning on triple coincidences while taking advantage of a special condition of spectrally uncorrelated photons \cite{mosley_08_heralded} (thus eliminating the need for spectral filtering in the trigger channel and dramatically increasing the heralding rate) recently led to the generation of the three-photon Fock state $ \vert 3 \rangle $ \cite{cooper_13_experimental}.

\subsection{Photon subtraction}

A single beam-splitter of low reflectivity and a photodetector are sufficient to implement single-photon subtraction from a light state.
\begin{figure}[h]
	\centering
	\includegraphics[width=7 cm]{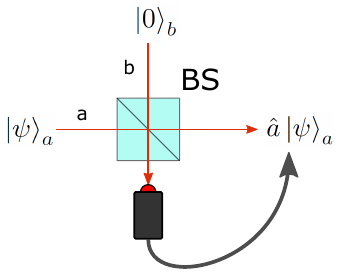}
	\caption{Scheme for heralded single-photon subtraction. A click from the on/off detector in the output mode $b$ heralds the application of the annihilation operator $\hat{a}$ on the input state $\vert \psi \rangle _a$. \label{fig_sub}}
\end{figure}
As shown in Fig.\ref{fig_sub}, if nothing is injected in the ancillary mode $ b $ and a single photon is detected at the output of the same mode, the operation is simply described by
\begin{equation}
	_b \langle 1 \vert \hat{B}(\tau) \vert \psi \rangle _a \vert 0 \rangle _b \approx \; _b\langle 1  \vert 1 + \tau (\hat{a}\hat{b}^{\dagger}-\hat{a}^{\dagger}\hat{b}) \vert \psi \rangle _a \vert 0 \rangle _b \rightarrow \hat{a}\vert \psi \rangle _a,
	\label{eq_sub}
\end{equation}
after normalization. This conditional operation thus corresponds to the action of the annihilation operator $ \hat{a} $ on a generic input state.
Of course, for Eq.(\ref{eq_sub}) to be valid, it is essential that the BS reflectivity is very low, otherwise higher-order terms in the Taylor expansion start to contribute, and the operation is no longer a simple single-photon annihilation. This is a particularly important condition since one normally uses avalanche photodiodes (APDs) that are not able to count the number of reflected photons. Therefore, limiting the BS reflectivity makes sure that the probability of having subtracted more than a single photon remains negligible.

Single-photon subtraction was first demonstrated by Wenger et al. \cite{wenger_04_nongaussian} to convert a Gaussian squeezed state into a highly non-Gaussian one. Later, photon subtraction was applied to a set of paradigmatic light states (a single photon, a coherent state, and a thermal one) \cite{zavatta_08_subtracting} to demonstrate its de-Gaussification effects and the peculiar behavior of photon subtraction on the mean photon number of the states. 

Single or multiple photon subtractions have also been extensively used as a resource for generating small Schr\"{o}dinger’s cats in pulsed or CW regimes \cite{ourjoumtsev_06_generating,neergaard-nielsen_06_generation,wakui_07_photon}. In optics, a Schr\"{o}dinger’s cat state \cite{schrodinger_35_gegenwartige}, which is a quantum superposition of macroscopically distinct states, is usually defined as the superposition
\begin{equation}\label{eq_cat}
	\vert cat_{\pm}\rangle= c (\vert \alpha \rangle \pm \vert -\alpha \rangle )
\end{equation}
of two coherent states of sufficiently large amplitude $ \vert \alpha \vert $ and opposite phase. Recalling the expansion of a coherent state into its Fock components, it is easy to see that the superposition with the + sign (a so-called 'even' cat) only contains even Fock terms, while the - sign superposition (an 'odd' cat) is only made of odd Fock components.
The starting point for these experiments and for the original subtraction experiment of \cite{wenger_04_nongaussian} was a so-called squeezed vacuum state, obtained by applying a single-mode squeezing operation
\begin{equation}
	\hat{S}_1(\zeta)=\exp \left[ \zeta ({\hat{a}^{\dagger 2}}-\hat{a}^2)\right]
	\label{eq_squeezer}
\end{equation}
(the same as Eq.(\ref{eq_pdc}), but acting on a single mode characterized by the annihilation operator $ \hat{a} $ instead of two modes with  $ \hat{a} $ and  $ \hat{b} $) to the vacuum state:
\begin{equation}\label{eq_sqv}
	\vert \psi _{sqv}\rangle=\hat{S}_1(\zeta) \vert 0 \rangle = \sqrt[4]{1- \lambda^2}\sum_{n=0}^{\infty} \frac{\sqrt{2n!}}{n!} (-\frac{1}{2}\lambda)^n \vert 2n \rangle .
\end{equation}
The squeezed vacuum state is clearly seen to be made of only even Fock terms and it is thus similar, at least for small amplitudes, to an even cat state $ \vert cat_{+}\rangle $. By subtracting a photon from such a state, a sum of only odd Fock terms is obtained, which faithfully simulates a small odd cat state $ \vert cat_{-}\rangle $ \cite{dakna_97_generating}. 
Recently, several other approaches have been used to generate or 'breed' optical Schr\"{o}dinger's cat superposition states, often using the particular result of a quadrature measurement (see Eq.(\ref{eq_quad})) on one mode of an entangled two-mode state to herald the production or the amplification of a cat state on the other \cite{ourjoumtsev_07_generation,laghaout_13_amplification,etesse_15_experimental,ulanov_16_losstolerant,sychev_17_enlargement}.

Different schemes have been recently introduced also for performing photon subtraction that are not based on a simple low-reflectivity BS. One of them relies on generating the sum-frequency signal from the input state and an additional intense gate pulse in a nonlinear crystal \cite{eckstein_11_quantum,ra_17_tomography}. When a photon is detected at the up-converted frequency, a single-photon subtraction has taken place. Differently from the BS scheme, this subtraction process is now mode-selective, since the mode of the subtracted photon is controlled by the shape of the gate pulse.
Another kind of photon subtraction is based on the use of phase modulation as a frequency BS to subtract a photon from a double sideband mode of continuous-wave light and thus generate an optical cat state \cite{serikawa_18_generation}.

In all the above cases, single-mode photon subtraction by the operator $\hat{a}$ is seen to be capable of de-Gaussifying the input states or enhancing their nonclassicality, but such an operation can never produce nonclassical behavior if it was not already present in the state since the beginning. It is worth reminding that the broadly accepted definition of nonclassicality relies on the Glauber-Sudarshan P-function \cite{glauber_63_coherent,sudarshan_63_equivalence}, and defines a state as nonclassical whenever it cannot be written as a statistical mixture of coherent states and its P-function is therefore not a proper probability distribution.

\subsection{Photon addition}

\begin{figure}[h]
	\centering
	\includegraphics[width=7 cm]{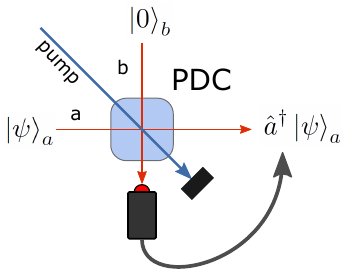}
	\caption{Scheme for heralded single-photon addition. A click from the on/off detector in the output mode $b$ heralds the application of the creation operator $\hat{a}^{\dagger}$ on the input state $\vert \psi \rangle _a$.\label{fig_add}}
\end{figure}

The same BS setup as discussed above could in principle be used to implement single-photon addition on an arbitrary input state by sending a single photon in the input ancillary mode and detecting the vacuum at its output \cite{nunn_21_heraldinga}. However, since all losses and detector inefficiencies would falsely trigger the heralding, thus mixing the desired operation with the identity, a better way of implementing the photon creation operator $ \hat{a} ^{\dagger}$ is by means of a parametric down-converter, as shown in Fig.\ref{fig_add}.

By sending an arbitrary light state in one input mode and vacuum on the other, the observation of a single-photon on the ancillary output mode $ b $ can be described as
\begin{equation}
	_b \langle 1 \vert \hat{S}(\zeta) \vert \psi \rangle _a \vert 0 \rangle _b \approx \; _b\langle 1  \vert 1 + \zeta (\hat{a}^{\dagger}\hat{b}^{\dagger}-\hat{a}\hat{b}) \vert \psi \rangle _a \vert 0 \rangle _b \rightarrow \hat{a}^{\dagger}\vert \psi \rangle _a,
	\label{eq_add}
\end{equation}
after normalization. Again, one has to assume small values of the squeezing factor $\zeta $, so that the first-order Taylor expansion of the exponential squeezing operator can be used and the probability of adding more than a single photon remains negligible. 

This technique was first used to add a photon to a coherent state \cite{zavatta_04_quantumtoclassical,zavatta_05_singlephoton} and later extended to arbitrary input ones \cite{zavatta_07_experimental}.
Contrary to the process of photon subtraction, the simple application of the creation operator to an arbitrary state of light turns it into a nonclassical state \cite{rahimi-keshari_13_quantum}, thanks to the removal of the vacuum component in its density matrix expressed in the basis of Fock states \cite{lee_95_theorem}. Remarkably, the degree of nonclassicality of the final state depends on the size of the initial one and can thus be finely adjusted.

It is worth noting that, due to its highly singular behavior, in general the Glauber-Sudarshan P-function of a state cannot be directly reconstructed from measured quantities. Since nonclassicality cannot be directly tested against its definition, various different approaches have been devised \cite{richter_02_nonclassicality,asboth_05_computable,park_21_verifying,biagi_21_experimental}, but the detection and quantification of the nonclassical character of a quantum state is still a much debated problem. Therefore, the availability of light states that can be continuously tuned across the border between the classical and the quantum domain is of extreme importance \cite{zavatta_07_experimental,kiesel_08_experimental}.

Recent proposals have suggested extending the concept of spectrotemporal-mode-selective non-Gaussian operations introduced earlier for photon subtraction, also to the case of photon addition. A theoretical analysis has shown under which experimental conditions it is possible to arbitrarily choose the unique mode in which the photon is added \cite{roeland_21_modeselective}.

Individual operations of photon addition and subtraction are extraordinary tools for manipulating light at the deepest level \cite{bellini_10_manipulating,kim_08_recent} and have already allowed reaching an impressive degree of control in the engineering of the quantum state of light. However, an even richer possibility of state manipulation can be achieved by properly combining them with other operations or among themselves.

\subsection{Displaced photon arithmetics: superpositions with the identity operator}

As early as 1999, Dakna et al \cite{dakna_99_generation} suggested that any arbitrary superposition of the first $ n $ Fock states of the form of Eq.(\ref{eq_fock_sup}) can be generated by applying a sequence of displaced photon addition operations
\begin{equation}
	\hat{a} ^{\dagger}+\gamma  \hat{\mathbbm{1}},
	\label{eq_displ_add}
\end{equation}
coherent superpositions of the photon creation operator and the identity operator $\hat{\mathbbm{1}}$, to the vacuum state. A few years later, Fiurasek et al.\cite{fiurasek_05_conditional} showed that the same goal can be achieved starting from a squeezed vacuum state and applying sequences of linear combinations of the photon subtraction and identity operators of the type
\begin{equation}
	\hat{a} +\gamma \hat{\mathbbm{1}}.
	\label{eq_displ_sub}
\end{equation}

\begin{figure}[h]
	\centering
	\includegraphics[width=13 cm]{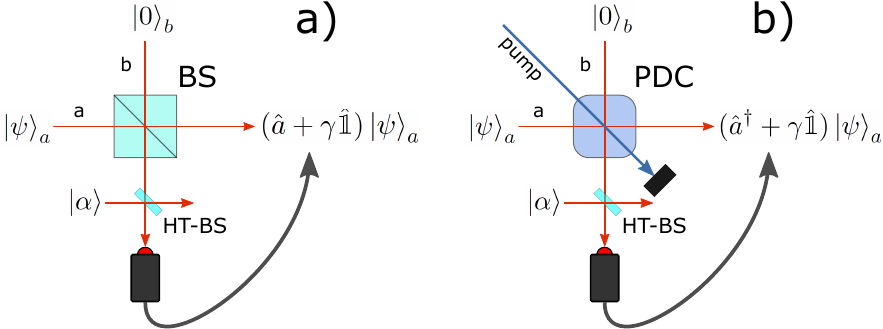}
	\caption{Schemes for displaced photon subtraction (a) and addition (b). If the herald mode $b$ is mixed with an ancilla coherent state $\vert \alpha \rangle$ on a high-transmittivity beam-splitter (HT-BS) before the on/off detector, a click heralds the coherent superposition of the addition or subtraction operation with the identity operator $\hat{I}$. \label{fig_displsubadd}}
\end{figure}

A simple way of experimentally implementing such linear superpositions of the photon addition or subtraction operations with the identity consists in modifying the basic setups described earlier by inserting an additional high-transmittivity beam-splitter (HT-BS), defined in Eq.(\ref{eq_bs_approx}), in the herald mode and injecting an ancilla coherent state in the other input, as shown in Fig.\ref{fig_displsubadd} (note that mixing an arbitrary optical quantum state with a coherent state in a high-transmittivity BS is the simplest way to experimentally implement the phase-space displacement operator $\hat D(\alpha)=e^{\alpha \hat{a}^{\dagger}-\alpha^* \hat{a}}$ \cite{paris_96_displacement}).

When a single photon is detected by the heralding APD, it is not in principle possible to determine its origin, whether from the subtraction BS (addition PDC crystal) or from the ancilla coherent state. In the first case, the simple subtraction (addition) operation took place; in the second case, nothing changed for the input states, so the identity operation was applied. In general, an arbitrary coherent superposition such as those described by Eqs.(\ref{eq_displ_add}) and (\ref{eq_displ_sub}) can thus be obtained by properly adjusting the HT-BS reflectivity and the coherent state complex amplitude $ \alpha $. 

Many variations of the above schemes have been used for generating a wealth of custom-made quantum states.
Simple coherent superpositions of vacuum and single-photon states can be produced by using the scheme depicted in Fig.\ref{fig_displsubadd}b) starting from a vacuum input state in mode $a$, as easily seen by applying the displaced addition operation of Eq.(\ref{eq_displ_add}) to $ \vert 0 \rangle $ \cite{resch_02_quantum}. Equivalently, one can obtain the same result starting from a single-photon state $ \vert 1 \rangle $ and applying the displaced photon subtraction of Eq.(\ref{eq_displ_sub}), shown in Fig.\ref{fig_displsubadd}a). 
 
The latter case is closely related to the so-called 'quantum scissors' scheme, first suggested by Pegg et al. \cite{pegg_98_optical}, which derives its name from the fact that, if both the beam-splitters have 50$ \% $ transmittivity, the output state is just a truncated version of the ancillary coherent state \cite{babichev_03_quantum,lvovsky_02_quantumoptical}.

If the setup for displaced photon subtraction of Fig.\ref{fig_displsubadd}a) is applied to a single-mode squeezed vacuum state $  \hat{S}_1(\zeta) \vert 0 \rangle  $ in mode $a$, the output is an arbitrary superposition of the original state and of its photon-subtracted version, equivalent to a squeezed single photon $  \hat{S}_1(\zeta) \vert 1 \rangle  $ \cite{takeoka_07_conditional,neergaard-nielsen_10_optical} (or, as we have seen earlier, an arbitrary superposition of small even and odd Schr\"{o}dinger's cats).  Each of the two states of the superposition is made of several Fock state components containing only even photon numbers for $  \hat{S}_1(\zeta) \vert 0 \rangle  $ and odd numbers for $  \hat{S}_1(\zeta) \vert 1 \rangle  $; they are thus orthogonal to each other, and together constitute a qubit basis of high interest for quantum information processing applications \cite{cochrane_99_macroscopically,ralph_03_quantum}.

Experimentally applying several operations of the kind of Eqs.(\ref{eq_displ_add}) and (\ref{eq_displ_sub}) in a sequential way is not straightforward, due to the growth in complexity, the accumulation of losses, and the drop of success probability over multiple steps. An alternative and related approach consists in using the two-mode squeezed vacuum state of Eq.(\ref{eq_epr}) and increasing the complexity of the operations performed in the heralding mode before detection. If the idler mode of the PDC is mixed with $ k $ weak ancillary coherent states on $ k $ additional symmetric BSs (in a configuration that combines the schemes of Figs.\ref{fig_fock} and \ref{fig_displsubadd}b), different combinations of clicks in APDs placed at their outputs herald different superpositions of Fock states up to $ k $ photons. Arbitrary coherent superpositions of Fock states up to two \cite{bimbard_10_quantumoptical} and three-photon states \cite{yukawa_13_generating,yukawa_13_emulating} have been produced this way, with the coefficients of the superposition depending on the amplitudes and phases of the ancillary coherent states.

\subsection{Sequences of photon additions and subtractions}

Besides applying the two basic operations of photon addition and subtraction independently onto single-mode light states, one may extend the range of experimentally viable state-manipulation techniques by concatenating such heralded operations in different sequences. A given sequence of clicks from the different detectors placed in the herald modes implements the corresponding operator sequence on the input state.

A sequence of two subtraction modules based on low-reflectivity beam-splitters, was implemented by Zavatta et al. \cite{zavatta_08_subtracting} for testing the effects of the photon annihilation operator on various light states. In particular,  the highly counter-intuitive increase of the mean photon number resulting from successive photon subtractions from a thermal state was experimentally verified.

\subsubsection{Noiseless amplification}
Concatenating two modules of different kind, Zavatta et al. \cite{zavatta_11_highfidelity} later realized a heralded amplifier for low amplitude coherent states characterized by a high fidelity. Its functioning is simply understood by examining the effect of the sequence of photon addition and subtraction $ \hat{a} \hat{a}^{\dagger} $ on a coherent state \cite{marek_10_coherentstate}:
\begin{equation}
\hat{a} \hat{a}^{\dagger} \vert \alpha \rangle \approx \hat{a} \hat{a}^{\dagger} (\vert 0 \rangle + \alpha \vert 1 \rangle + ...)=\vert 0 \rangle + 2\alpha \vert 1 \rangle + ...)\approx  \vert 2\alpha \rangle
\label{eq_nla}
\end{equation}
i.e., the final state is a coherent state of double amplitude. Doubling the state amplitude while preserving the width of its quadrature distributions reduces the phase variance roughly by a factor of 4.
This is an example of phase-insensitive noiseless linear amplification, a process that is normally forbidden in quantum mechanics, at least in a deterministic fashion. Were it possible, the transformed bosonic operators would no longer obey the commutation relations \cite{caves_82_quantum}. Phase-insensitive amplification is thus always accompanied by additional noise of quantum origin and cannot be used for cloning quantum states \cite{wootters_82_single}, improving phase sensitivity, or for state discrimination.
One can overcome this limitation recurring to a non-deterministic linear amplifier, as first suggested by Ralph and Lund in 2008 \cite{ralph_09_nondeterministic}. So, instead of the usual deterministic noisy amplification, probabilistic noiseless amplification schemes may be devised \cite{barbieri_11_nondeterministic}.
The first proposal was based on a quantum scissors scheme as described in the previous section, and was soon experimentally realized to amplify a small coherent state in a truncated Hilbert space of 0 and 1 photons \cite{ferreyrol_10_implementation,xiang_10_heraldeda}. However, due to this abrupt state truncation, the amplified states had a relatively low fidelity to coherent states and an effective amplification was only demonstrated for input states of very small amplitude. Non-deterministic noiseless amplification may find applications in different fields, from its inclusion in quantum communication schemes \cite{gisin_10_proposal,bruno_13_complete,kocsis_13_heralded}, as an entanglement distillation device \cite{xiang_10_heraldeda,he_21_noiseless}, or for improved phase estimation \cite{usuga_10_noisepowered}. 

\subsubsection{Test of non-commutativity}
By placing a module for single-photon creation between two modules for single-photon annihilation and choosing the right combination of the clicks coming from the module detectors, a range of different sequences of photon additions and subtractions can be implemented. In particular, by heralding with coincidences between the addition module and one of the subtraction ones, one can either produce a first-subtracted-then-added state or vice versa, depending on the order of clicks.
In 2007, Parigi et al. \cite{parigi_07_probinga} applied the two inverse sequences $ \hat{a} \hat{a}^{\dagger} $ and $ \hat{a}^{\dagger} \hat{a} $ to thermal light states in order to test the bosonic commutation rules. The two final states were analyzed by homodyne detection and found to be completely different from each other and from the original thermal one, thus providing the first direct experimental verification of the noncommutativity of the quantum bosonic creation and annihilation operators.

\subsection{Superpositions of additions and subtractions}
\begin{figure}[h]
	\centering
	\includegraphics[width=12 cm]{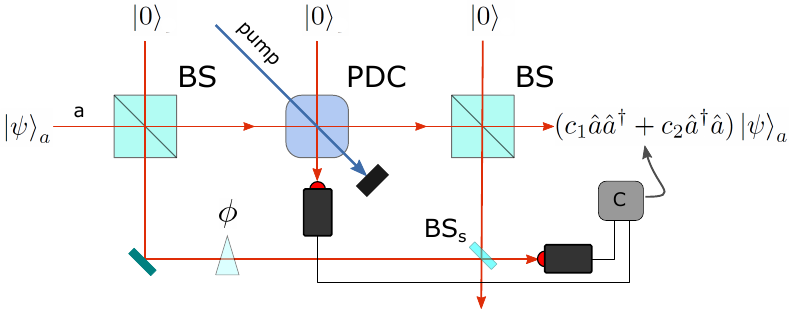}
	\caption{Scheme for the coherent superposition of the two inverse sequences of photon addition and subtraction $\hat a \hat a^{\dagger}$ and $\hat a^{\dagger} \hat a$. One module for photon addition is placed between two modules for photon subtraction and the desired operation is heralded by the coincident click from the addition detector and from another detector placed after the mixing (with an adjustable relative phase $\phi$) of the two subtraction herald modes on a beam-splitter (BS$_s$). \label{fig_supscheme}}
\end{figure}

Implementing coherent superpositions of heralded quantum operations relies on the indistinguishability of the herald photons coming from different processes. It may be obtained, for example, by mixing them on a beam-splitter before detection, in a way analogous to the case of the superpositions with the identity operator described earlier.
A particularly interesting experimental configuration uses the three-module scheme described above and mixes the herald photons from the two photon subtraction stages in a beam-splitter before coincident detection with the herald photon coming from the addition stage, as shown in Fig.\ref{fig_supscheme} \cite{kim_08_scheme}.

This results in the indistinguishability and coherent superposition of the two inverse sequences of operators $\hat a \hat a^{\dagger}$ and $\hat a^{\dagger} \hat a$. By playing with the relative phase between the herald modes of the two subtraction modules and with their mixing ratio in the beam-splitter, arbitrary superpositions of the two sequences of the kind
\begin{equation}\label{eq_sup_addsub}
	c_1 \hat a \hat a^{\dagger} + c_2\hat a^{\dagger} \hat a 
\end{equation}
may be realized.

\subsubsection{Test of bosonic commutation rules}
Such an approach was first used, in the case of photon addition and subtractions, to experimentally implement the commutator of bosonic creation and annihilation operators \cite{kim_08_scheme,zavatta_09_experimental}. In particular, the commutator $ \left[ \hat a, \hat a^{\dagger}\right]= \hat a \hat a^{\dagger}-\hat a^{\dagger} \hat a $ was realized by mixing the herald photons on a balanced 50$ \% $ BS while setting to 0 the relative phases. The state resulting from the application of the commutator was found to be identical to the input one, thus verifying that such an operation is equivalent to the identity. On the other hand, the anti-commutator, realized with a relative phase pf $\pi$, implements the operation $ 2 \hat a ^\dag \hat a + K \hat 1 $, which generates a state that is strongly dependent on the value of the constant $K$. From the homodyne measurements of the generated state, it was possible to quantitatively demonstrate the bosonic relation by measuring the value of $K$.

\subsubsection{Generalized noiseless amplification and emulation of Kerr nonlinearities}
Fiurasek \cite{fiurasek_09_engineering} found that the basic scheme of Fig.\ref{fig_supscheme} may be generalized (by including photon-number-resolving detectors able to count up to $ N $ photons instead of the APDs of Fig.\ref{fig_supscheme}) for the approximate probabilistic realization of arbitrary operations $ f(\hat{n}) $ that can be expressed as a function of photon number operator $ \hat{n}=\hat a^{\dagger} \hat a $. This class of transformations includes for instance the general noiseless linear amplifier $\hat{Z}=g^{\hat{n}} $, where $ g>1 $ is the amplification gain, or the Kerr nonlinearity, described by a unitary operation $ \hat{U}=\exp (-i \phi \hat{n}^2) $.

In the simple case of $ N=1 $, the noiseless linear amplifier $ \hat{Z} $ operator may be approximated by a first-order polynomial in $ \hat{n} $ that reads: 
\begin{equation}\label{eq_nla_gen}
	\hat{Z}_1=(g-1) \hat{n}+1=(g-2) \hat a^{\dagger} \hat a + \hat a \hat a^{\dagger}
\end{equation}
where the commutation relation has been used in the right hand side.
Such a variable-gain noiseless amplifier is readily realizable with the setup of Fig.\ref{fig_supscheme} and it is easy to see that the particular case of $ g=2 $ corresponds to the noiseless amplifier based on the simple $ \hat a \hat a^{\dagger} $ sequence described earlier \cite{zavatta_11_highfidelity}.

The emulation of a Kerr nonlinearity is more resource-demanding than noiseless amplification because already the first nontrivial approximation would require subtraction and addition of two photons, $ \hat{U}_2=1 +i \phi \hat{n}^2 $. However, it can still be approximately implemented making use of Eq.(\ref{eq_sup_addsub}) if one targets a well-defined nonlinearity that induces a $ \pi$-phase shift of the two-photon Fock component $ \vert 2 \rangle $ of the state with respect to $ \vert 0 \rangle $ and $ \vert 1 \rangle $. In the experiment by Costanzo et al. \cite{costanzo_17_measurementinduced}, this interaction was successfully emulated with the basic scheme of Fig.\ref{fig_supscheme} on the smallest nontrivial subspace spanned by the vacuum, single-photon, and two-photon Fock components of a weak input coherent state. 

\subsubsection{State orthogonalization and CV qubit generation}
A universal quantum NOT operation, which brings an arbitrary state to its orthogonal one, is not possible without some prior knowledge of the input state \cite{buzek_99_optimal}, just like it is impossible to perfectly and deterministically clone or amplify a quantum state \cite{wootters_82_single,caves_82_quantum} without prior information. However, in 2013 Vanner et al.\cite{vanner_13_quantum} suggested that a perfect orthogonalizer can be in principle realized even if only some very limited preliminary information about the input state is available. This idea was later generalized by Coelho et al. \cite{coelho_16_universala}, who showed that, given an arbitrary operator $ \hat{C} $, if one knows its mean value $ \langle \hat{C} \rangle $ for the input state $ \vert \psi \rangle $, the operation
\begin{equation}\label{eq_ortho}
	\hat{O}_C \equiv \hat{C} - \langle \hat{C} \rangle \hat{\mathbbm{1}}
\end{equation}
converts the original state to an orthogonal one. Moreover, a simple change in the coefficient of the identity operator in the above expression allows one to produce any coherent superposition of the original input state $ \vert \psi \rangle $ and of its orthogonal state $ \vert \psi_{\perp} \rangle $, which constitutes the most general form of an arbitrary CV qubit.
In the experiment, Coelho et al. \cite{coelho_16_universala} realized such an operation in two different ways: the first used the photon creation operator and implemented Eq.(\ref{eq_ortho}) with $ \hat{C} \equiv \hat{a}^{\dagger}  $. The coherent superposition of photon addition and identity operator was thus realized with a setup based on Fig.\ref{fig_displsubadd}b).
The second scheme used the photon number operator, letting $ \hat{C} \equiv \hat{n} $ in Eq.(\ref{eq_ortho}), and was realized by properly adjusting the superposition coefficients of Eq.(\ref{eq_sup_addsub}) in the generic experimental scheme of Fig.\ref{fig_supscheme}.

\section{Multimode state engineering by delocalized addition and subtraction}

The superpositions of heralded operations described in the previous section always took place on the same traveling mode of light. If two or more conditional operations act on different field modes and their herald photons are mixed and made indistinguishable before detection, a more general multimode coherent superposition of quantum operations can be realized.
\begin{figure}[h]
	\centering
	\includegraphics[width=10 cm]{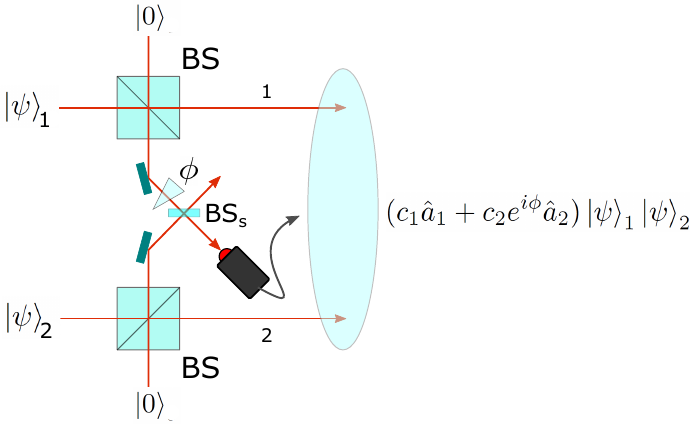}
	\caption{Scheme for coherent photon subtraction from two distinct spatial modes, 1 and 2. \label{fig_mmsub}}
\end{figure}
Figure \ref{fig_mmsub} shows an ideal experimental scheme for applying a coherent superposition of photon subtractions to two different spatial modes, labeled 1 and 2. The reflected modes of two low-reflectivity beam-splitters acting separately on each mode  are mixed in a further beam-splitter (BS$_s$) after some additional phase retardation $ \phi $ to realize a delocalized photon subtraction operation of the kind:
\begin{equation}\label{eq_delocsub}
	c_1 \hat{a}_1 + c_2 e^{i \phi}\hat{a}_2 .
\end{equation}

In the scheme of Fig.\ref{fig_mmadd} it is the idler modes of two PDC crystals that are mixed in the beam-splitter BS$_s$ to realize the delocalized photon addition operation
\begin{equation}\label{eq_delocadd}
	c_1 \hat{a}^{\dagger}_1 + c_2 e^{i \phi}\hat{a}^{\dagger}_2 .
\end{equation}

Of course, the above ideal two-mode schemes can be readily generalized to the coherent addition and subtraction of single photons to and from an arbitrary number of distinct field modes. Moreover, such modes need not necessarily be different spatial ones, as those of Figures \ref{fig_mmsub} and \ref{fig_mmadd}, but they may in principle involve any other degree of freedom of light.
In the following subsections we will illustrate examples of these coherent multimode operations and their impact upon the generation and enhancement of entanglement. Most of the experimental results obtained so far have been achieved in two-mode configurations involving either two distinct spatial or temporal wavepacket modes. However, the field is rapidly evolving and, especially thanks to the inherent compactness and phase stability of spectrotemporal degrees of freedom, is already pushing experimental research towards larger multimodal configurations.

\subsection{Multimode photon subtraction and entanglement distillation}

Entanglement is a purely quantum feature that, besides its fundamental relevance, constitutes an important resource for quantum information processing, at the base of many protocols that have no counterpart in the classical world, like teleportation \cite{boschi_98_experimental,furusawa_98_unconditional} and secure communication \cite{ekert_91_quantum}.
In optics, entangled states are usually generated by a PDC process in the form of a two-mode squeezed vacuum (or EPR state), as in Eq.(\ref{eq_epr}), which is represented by a Gaussian function in phase space. Since entangled states are extremely fragile and quickly deteriorate with losses, one would like to recover their entanglement by some distillation procedure. However, it is well-known that the entanglement of a Gaussian state cannot be improved by ordinary linear optical devices that implement Gaussian operations \cite{eisert_02_distilling}.
Entanglement distillation of CV Gaussian states thus requires non-Gaussian operations, and photon subtraction is probably the most straightforward way to achieve it, as first proposed by Opatrný et al. \cite{opatrny_00_improvement} and recently further discussed in \cite{zhang_22_maximal}. Experiments have shown that local photon subtractions in one or both modes of a two-mode squeezed vacuum state indeed led to an increase of the degree of entanglement \cite{takahashi_10_entanglement,kurochkin_14_distillation}.
However, a more efficient distillation scheme, involving the delocalized subtraction of a single photon described by Eq.(\ref{eq_delocsub}) from the two modes of an EPR state, was demonstrated by Ourjoumtsev et al. in 2007 \cite{ourjoumtsev_07_increasing}. 
Recently, Maga\~{n}a-Loaiza et al. \cite{magana-loaiza_19_multiphoton} utilized a bright SPDC source in combination with photon-number-resolving detectors, to demonstrate the generation of a family of correlated photon-subtracted two-mode squeezed vacua with a broad range of mean photon numbers and degrees of correlation.

The possibility of subtracting a single photon from an arbitrary spectrotemporal mode by mode-selective frequency up-conversion \cite{ra_17_tomography} can also be extended to the subtraction from a coherent superposition of multiple time-frequency modes \cite{averchenko_14_nonlinear,averchenko_16_multimode}. Starting from a Gaussian multimode squeezed vacuum state, recent experiments have demonstrated the heralded generation of states with negative Wigner functions in controlled modes and in different superpositions of them, thus producing non-Gaussian entanglement \cite{ra_20_nongaussian}.

\subsection{Multimode photon addition and entanglement generation}
Just as  the operation of adding a photon to a light field in a well-defined single mode has had such a dramatic impact on extending the possibilities of photonic quantum technologies, in the same way, the ability to coherently add a single photon to two or more different light modes by a general superposition $\sum_m c_m \hat a^{\dag}_m$ (where the subscript $m$ indexes the different modes and the $c_m$ are complex coefficients) may open new avenues in multimode quantum state manipulation and control.
One common feature of the application of a superposition of photon addition operations to multimode states of light is that entanglement is invariably generated at the output, independently on the input. This is in striking contrast with the effect of coherent photon subtractions which, as shown above, can enhance the nonclassicality and distill the entanglement already present among the modes, but cannot generate it from scratch.
\begin{figure}[h]
	\centering
	\includegraphics[width=10 cm]{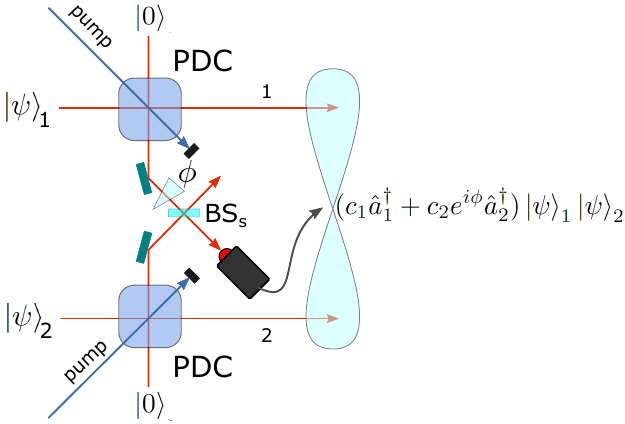}
	\caption{Scheme for coherent photon addition to two distinct spatial modes, 1 and 2. \label{fig_mmadd}}
\end{figure}

The coherent addition of a single photon to multiple modes is thus the simplest way to generate entanglement among modes containing arbitrary, even classical and uncorrelated, states of light. It has already proved itself a powerful tool for studying fundamental quantum physics,  and one may expect it to become an invaluable resource also for the development of future quantum-enhanced technologies \cite{biagi_21_coherent}.

The earliest and simplest example of delocalized single photon addition in a CV setting was realized in 2006 \cite{zavatta_06_remote}  and involved two distinct temporal wavepacket modes initially empty. The temporal-mode version of the basic scheme of Fig.\ref{fig_mmadd} was applied to the input quantum state $\vert 0\rangle_1 \vert 0\rangle_2$, producing the so-called single-photon mode-entangled state
\begin{equation}
	\vert \psi_{SP}\rangle_{12} = c_1\vert 1\rangle_1 \vert 0\rangle_2 + e^{i\phi} c_2\vert 0\rangle_1 \vert 1\rangle_2,
	\label{eq_spdeloc}
\end{equation}
which is normalized under the condition $c_1^2+c_2^2=1$. Note that such a CV scheme is functionally equivalent to the source of polarization-entangled photons first introduced in \cite{kwiat_99_ultrabright} and traditionally used in DV experiments.

Since its first demonstration, this experimental approach has been used in several different contexts and with different combinations of light states in the input modes. 

Applying a coherent superposition of single-photon addition operations onto two distinct modes containing vacuum and a coherent state, $\vert 0\rangle_1 \vert \alpha \rangle_2$, resulted in the generation of a so-called hybrid entangled states \cite{jeong_14_generation,costanzo_15_properties}. These states present entanglement between two distinct modes where light is best described in the discrete- (DV) or continuous-variable (CV) type of encoding, respectively. Single photons are typical examples of DV states,  and typical DV qubits can be made of superpositions of the presence and absence of a single photon in a particular field mode. Conversely, coherent states of light are typical examples of CV states \cite{weedbrook_12_gaussian}, and CV qubits can be made of the superposition of two coherent states of different complex amplitude (although recent attention has partly shifted towards so-called Gottesman-Knill-Preskill (GKP)  states \cite{gottesman_01_encoding} that have been shown to allow for the correction of arbitrary types of noise).
Since a single-photon-added coherent state \cite{agarwal_91_nonclassical,zavatta_04_quantumtoclassical} resulting from the application of the photon creation operator to a coherent state $\vert \alpha \rangle$, can well approximate another coherent state of slightly larger amplitude $\vert \alpha' \rangle$ whenever $|\alpha|$ is sufficiently large, applying the generic superposition of Eq.(\ref{eq_delocadd}) with appropriate coefficients may result in the production of the hybrid DV-CV entangled state
\begin{equation}
	\vert \psi_H\rangle_{12} = \frac{1}{\sqrt{2}} (\vert 1\rangle_1 \vert \alpha\rangle_2 + e^{i\phi}\vert 0\rangle_1 \vert\alpha'\rangle_2).
	\label{eq_hybrid}
\end{equation}
A different but closely related form of hybrid CV-DV entangled state was proposed in \cite{andersen_13_heralded} and experimentally realized simultaneously with the coherent photon addition scheme described above by Morin et al. \cite{morin_14_remotea}. It used the indistinguishability of the origin of a herald photon coming either from the idler mode of a two-mode squeezed vacuum, or from subtraction from a single-mode squeezed state. Recent works by the same group have further investigated the properties of such a hybrid entangled state and its possible applications in heterogeneous quantum networks \cite{huang_19_engineering,lejeannic_18_remote,guccione_20_connecting}.

The effect of a coherent superposition of photon addition operations was finally experimentally evaluated in the case of two input modes containing identical coherent states $\vert \alpha\rangle$~\cite{biagi_20_entangling}. 
The state produced by applying the balanced version (with $ c_1=c_2 $) of the operator superposition of Eq.(\ref{eq_delocadd}) on the input state $\vert \alpha\rangle_1 \vert \alpha\rangle_2$ can be written as:
\begin{eqnarray}
	\vert \psi_{\phi}(\alpha)\rangle_{12}
	&=&\Big[\hat D_1(\alpha)\hat D_2(\alpha) \Big( \vert 1\rangle_1 \vert 0\rangle_2 + e^{i \phi}\vert 0\rangle_1 \vert 1\rangle_2\Big) 
	\nonumber\\
	&+&\alpha^*(1+e^{i \phi})\vert \alpha\rangle_1 \vert\alpha\rangle_2\Big]/\sqrt{\mathcal{N}}
	\label{eq_displace}
\end{eqnarray}
with the normalization factor $\mathcal{N} = 2[1+|\alpha|^2(1+\cos\phi)]$. The output state is made of an entangled and a separable part, whose relative weights depend on the superposition phase $\phi$ and the amplitude $\alpha$ of the coherent states.
When $\phi=0$, the entanglement of the state decreases for increasing $\alpha$, due to the growing contribution of the separable fraction. When the other extreme condition of $\phi=\pi$ is reached, the separable part disappears and the resulting state reduces to a displaced delocalized single photon. Since displacing a state in phase space does not change its entanglement, one can preserve it at its constant maximum value independently of the amplitude of the input coherent states and even between two modes initially containing large mean photon numbers $\bar n =|\alpha|^2$.
Such a state, characterized by a degree of entanglement independent of the size of the entangled partners and surprisingly robust against losses \cite{sekatski_12_proposal}, is of very high interest to investigate the resilience and detectability of entanglement for states of growing macroscopicity  as well as  for fundamental investigations concerning the very definition of macroscopic quantumness and entanglement \cite{lee_11_quantification,frowis_18_macroscopic,jeong_15_characterizations,laghaout_15_assessments}. Related proposals and experiments involving micro--macro~\cite{andersen_13_heralded,bruno_13_displacement,lvovsky_13_observation} and macro--macro \cite{sychev_19_entanglement} entanglement have recently explored these issues, with different approaches.

The experiment by Biagi et al.\cite{biagi_20_entangling} demonstrated the preservation of a high degree of entanglement even for macroscopic mean photon numbers (up to $\bar n \approx 60$) in each mode, but also revealed its extreme sensitivity to small variations of the superposition phase $ \phi $, which might find applications for remote quantum sensing purposes \cite{biagi_21_remotea}.

The state resulting from the balanced superposition of photon additions on two modes containing identical coherent states and described by Eq.(\ref{eq_displace}) with $\phi=\pi$, was also found to show another interesting property, called discorrelation \cite{meyer-scott_17_discorrelated,biagi_21_generating} . In discorrelated states, the number of photons in each mode can take any value individually, but two modes together never exhibit the same. For a two-mode state, it manifests itself in null diagonal elements of the joint photon number probability distribution $P_{n_1,n_2}$, where $n_1$ and $n_2$ are the numbers of photons in the two modes, i.e., $P_{n,n}=0$, while the marginal distributions $P_{n_1}=\sum_{n_2=0}^{\infty}P_{n_1,n_2}\neq 0$, for any value of $n_1$. This property could be used to distribute unique randomness between parties, particularly in so-called `mental poker' problems, which are concerned with the fair dealing of cards between distant players without a trusted third party \cite{shamir_81_mental,rivest_78_method}.

\section{Conclusions}
In this review, we have concisely presented recent results in the quantum engineering of light states achieved by the accurate manipulation of the electromagnetic field at the single-photon level. This is an active and lively research field and new ideas and experiments appear every day from many groups all around the world.
By combining a few basic building blocks in different experimental configurations, complex operations can be implemented that transform the input light states into new ones, with accurately designed characteristics. Thanks to the many parameters that can be adjusted in the processes (such as the weights and relative phases of the operator superpositions, or the type of input states), a huge variety of quantum states can be generated, and their properties may be arbitrarily tuned in a controllable way. 
These powerful tools can enable both the verification of fundamental quantum laws of nature and the realization of novel devices for advanced quantum technologies, ranging from the communication and processing of quantum information, to quantum-enhanced sensing and metrology.

\bibliography{myrefs}

\end{document}